\documentclass[preprint,showpacs,preprintnumbers,amsmath,amssymb,showkeys]{revtex4}
 
\usepackage{graphicx}
\usepackage{epsfig}		
\usepackage{dcolumn}
\usepackage{bm}

\def\0{\mbox{\tiny $0$}}
\def\1{\mbox{\tiny $1$}}
\def\2{\mbox{\tiny $2$}}
\def\3{\mbox{\tiny $3$}}
\def\4{\mbox{\tiny $4$}}
\def\5{\mbox{\tiny $5$}}
\def\6{\mbox{\tiny $6$}} 
\def\7{\mbox{\tiny $7$}}
\def\8{\mbox{\tiny $8$}}
\def\9{\mbox{\tiny $9$}}
\def\R{\mbox{\tiny $R$}}
\def\T{\mbox{\tiny $T$}}

\def\bb{\mbox{\tiny $B$}}

\def\f14{\mbox{\tiny $\frac{1}{4}$}}

\def\L{\mbox{\tiny $L$}}

\begin{document}

\title{SMALL CORRECTIONS TO THE TUNNELING PHASE TIME FORMULATION}

\author{A. E. Bernardini}
\email{alexeb@ifi.unicamp.br}
\affiliation{Instituto de F\'{\i}sica Gleb Wataghin, UNICAMP,\\
PO Box 6165, 13083-970, Campinas, SP, Brasil.}

\date{\today}

\begin{abstract}
After reexamining the above barrier diffusion problem where we notice that the wave packet collision implies the existence of {\em multiple} reflected and transmitted wave packets, we analyze the way of obtaining phase times for tunneling/reflecting particles in a particular colliding configuration where the idea of multiple peak decomposition is recovered.
To partially overcome the analytical incongruities which frequently rise up when the stationary phase method is adopted for
computing the (tunneling) phase time expressions, we present a theoretical exercise involving a symmetrical collision between two identical wave packets and 
a unidimensional squared potential barrier where the scattered wave packets can be recomposed by summing the amplitudes of simultaneously reflected and transmitted wave components so that the conditions for applying the stationary phase principle are totally recovered.
Lessons concerning the use of the stationary phase method are drawn.
\end{abstract}








\maketitle

The recent developments of nanotechnology brought new urgency to study the tunneling time as it is directly
related to the maximum attainable speed of nanoscale electronic devices.
In parallel, a series of recent experimental results \cite{Nim92,Ste93,Chi98,Hay01}, some of them corroborating with the possibility of
superluminal tunneling speeds for photons, 
have revived an interest in the tunnelling time analysis \cite{Olk04,Pri03,Lan94,Olk92,Hau89}.
On the theoretical front, people have tried to introduce quantities that have the dimension
of time and can somehow be associated with the passage of the particle through the barrier or,
strictly speaking, with the definition of the tunneling time.
Since a long time these efforts have led to the introduction of several {\em time} definitions \cite{Olk04,Baz67,But83,Hau87,Fer90,Yuc92,Hag92,Bro94,Olk95,Jak98,Olk02}, some of which are
completely unrelated to the others, which can be organized into three groups:
(1) The first group comprises a time-dependent description in terms of wave packets where some features
of an incident wave packet and the comparable features of the transmitted packet are 
utilized to describe a {\em delay} as tunneling time \cite{Hau89}.
(2) In the second group the tunneling times are computed with basis on averages over a set of kinematical paths,
whose distribution is supposed to describe the particle motion
inside a barrier, i. e. Feynman paths are used like real paths
to calculate an average tunneling time with the weighting
function $\exp{[i\, S\, x(t)/\hbar]}$, where $S$ is the action associated with
the path $x(t)$ (where x(t) represents the Feynman paths initiated
from a point on the left of the barrier and ending at another
point on the right of it \cite{Sok87}).
The Wigner distribution paths \cite{Bro94}, and the Bohm approach \cite{Ima97,Abo00} are included in this group.
(3) In the third group we notice the introduction of a new degree of freedom,
constituting a physical clock for the measurements of tunnelling times.
Separately, it stands by itself the dwell time approach.
The time related to the tunneling process is defined by the interval during which the incident flux has to exist and act,
to provide the expected accumulated particle storage, inside the barrier \cite{Lan94}.
The methods with a Larmor clock \cite{But83} or an oscillating barrier \cite{But82} are comprised by this group.

There is no general agreement \cite{Olk04,Olk92} among the above definitions about the meaning of tunneling times 
(some of the proposed tunneling times are actually traversal times, while others seem to represent in reality only the
spread of their distributions) and about which, if any, of them
is the proper tunneling time, in particular, due to the following reasons \cite{Olk04}:
(a) the problem of defining tunnelling times is closely connected with the more general definition of the
quantum-collision duration, and therefore with the fundamental fact that 
in quantum mechanics, time enters as a parameter rather than an observable to which an operator can be assigned
(b) the motion of particles inside a potential barrier is a quantum phenomenon, that till now has been devoid of any
direct classical limit; 
(c) there are essential differences among the initial, boundary and external
conditions assumed within the various definitions proposed in the literature; those differences have
not been sufficiently analyzed yet. 
In particular, the study of tunneling mechanisms is embedded by theoretical constructions involving analytically-continuous {\em gaussian}, or infinite-bandwidth step pulses to examine the tunneling process.
Nevertheless, such holomorphic functions do not have a well-defined front in a manner that the interpretation of the wave packet speed of propagation becomes ambiguous.
Moreover, infinite bandwidth signals cannot propagate through any real physical medium (whose transfer function is
therefore finite) without pulse distortion, which also leads to ambiguities in determining the propagation velocity during the tunneling process.
For instance, some of the barrier traversal time definitions lead, in tunneling time conditions, to very short times which can even become negative, precipitately
inducing to an interpretation of violation of simple concepts of causality.
Otherwise, negative speeds do not seem to create problems with causality,
since they were predicted both within special relativity and within quantum mechanics \cite{Olk95}.
A possible explanation of the time advancements related to the negative speeds can come, in any case, from consideration of the very rapid
spreading of the initial and transmitted wave packets for large momentum distribution widths.
Due to the similarities between tunnelling (quantum) packets and evanescent (classical) waves,
exactly the same phenomena are to be expected in the case of classical barriers
(we can mention the analogy between
the stationary Helmholtz equation for an electromagnetic wave packet - in a waveguide, for instance - in
the presence of a {\em classical} barrier and the stationary Schroedinger equation, in the presence of a potential barrier \cite{Lan94,Jak98,Nim94}).
The existence of such negative times is predicted by relativity itself, on the basis of the ordinary postulates \cite{Olk04}, and they appear to have been experimentally detected in many
works \cite{Gar70,Chu82}.

In this extensively explored scenario,
the first group quoted above contains the so-called phase times \cite{Boh52,Wig55}
which are obtained when the stationary phase method (SPM) \cite{PBE}
is employed for obtaining the times related to the motion of the wave packet spatial centroid
which adopts averages over fluxes pointing in a well-defined direction only, and has recourse to a quantum operator
for time \cite{Olk04}.
Generically speaking, the SPM essentially enables us 
to parameterize some subtleties of several quantum phenomena,
such as tunneling \cite{Hau89,Ste93,Bro94},
resonances \cite{Bra70}, incidence-reflection
and incidence-transmission interferences \cite{Per01}
as well as the Hartman Effect \cite{Har62} and
its {\em superluminal} traversal time interpretation \cite{Olk04,Lan94,Jak98}.
In fact, it is the simplest and commonest approximation method for describing the group velocity of
a wave packet in a quantum scattering process represented by a collision of a particle with a potential barrier
\cite{Olk04,Lan94,Hau87,Wig55,Har62,Ber04}.

Our attention is particularly concentrated on some limitations on the use of the SPM for deriving tunneling times 
for which we furnish an accurate quantification of the analytical incongruities which restrict the applicability of this method.
We introduce a theoretical construction involving a symmetrical collision with a unidimensional square potential where the
scattered wave packets can be reconstructed by summing the amplitudes of the reflected and transmitted waves 
in the scope of what we denominate a multiple peak decomposition analysis \cite{Ber04} in a manner that
the analytical conditions for the SPM applicability are totally recovered.

Generically speaking, the SPM can be successfully utilized for describing
the movement of the center of a wave packet constructed in terms of a momentum distribution $g(k - k_{\0})$
which has a pronounced peak around $k_{\0}$. 
By assuming the phase which characterizes the propagation varies sufficiently smoothly
around the maximum of $g(k - k_{\0})$, the stationary phase condition enable us to calculate the position of the
peak of the wave packet (highest probability region to find the propagating particle).
With regard to the tunneling effect, the method is indiscriminately applied to find the position of a wave packet 
which traverse a potential barrier.
For the case we consider the potential barrier
\small\begin{equation}
V(x) = \left\{\begin{array}{cll} V_o && ~~~~x \in \mbox{$\left[- L/2, \, L/2\right]$}\\ &&\\ 0&& ~~~~x \in\hspace{-0.3cm}\slash\hspace{0.1cm}\mbox{$\left[- L/2, \, L/2\right]$}\end{array}\right.
\label{2p60}
\end{equation}
it is well known that the transmitted wave packet solution ($x \geq L/2 $) calculated by means of the Schroedinger formalism is 
given by \cite{Coh77}
\small\begin{equation}
\psi^{\T}(x,t) = \int_{_{\0}}^{^{w}}\frac{dk}{2\pi} \, g(k - k_{\0}) \, |T(k, L)|\,
\exp{\left[ i \, k \,(x - L/2) - i \, \frac{k^2}{2\,m} \, t +
 i \,\Theta(k, L)\right]},
\end{equation}\normalsize
where, in case of tunneling, the transmitted amplitude is written as
\small\begin{equation}
|T(k, L)| =
\left\{1+ \frac{w^4}{4 \, k^2 \, \rho^{\2}(k)}
\sinh^2{\left[\rho(k)\, L \right]}\right\}^{-\frac{1}{2}},
\label{1}
\end{equation}\normalsize
and the phase shift is obtained in terms of
\small\begin{equation}
\Theta(k, L) = \arctan{\left\{\frac{2\, k^2 - w^2}
{k \rho(k)}
\tanh{\left[\rho(k) \, L \right]}\right\}}, 
\label{502}
\end{equation}\normalsize
for which we have made explicit the dependence on the barrier length $L$ and 
we have adopted $\rho(k) = \left(w^2 - k^2\right)^{\frac{1}{2}}$ with $w = \left(2\, m \,V_{\0}\right)^{\frac{1}{2}}$ and $\hbar = 1$.
Without thinking over an eventual distortion that $|T(k, L)|$ causes to the supposedly symmetric function $g(k - k_{\0})$,
when the stationary phase condition is applied to the phase of Eq.~(\ref{1}), we obtain
\begin{eqnarray}
\frac{d}{dk}\left.\left\{k \,(x - L/2) - \frac{k^2}{2\,m} \, t 
+ \Theta(k, L)\right\}\right|_{_{k = k_{\mbox{\tiny max}}}} 
&=&0 ~~~~\Rightarrow\nonumber\\
  x - L/2 - \frac{k_{\mbox{\tiny max}}}{m} \, t + 
\left.\frac{d\Theta(k, L)}{dk}\right|_{_{k = k_{\mbox{\tiny max}}}} &=& 0.
\label{3}
\end{eqnarray}
The above result is frequently adopted for calculating the transit time $t_{T}$ of
a transmitted wave packet when its peak emerges at $x = L/2$,
\begin{eqnarray}\small
t_{T} &=&\frac{m}{k_{\mbox{\tiny max}}}\left.\frac{d\Theta(k, \alpha_{(\L)})}{dk}\right|_{_{k = k_{\mbox{\tiny max}}}}\nonumber\\
 &=&
\frac{2\,m \, L}{k_{\mbox{\tiny max}} \,\alpha }
\left\{\frac{w^4\,\sinh{(\alpha)}\cosh{(\alpha)}
-\left(2\, k_{\mbox{\tiny max}}^2 - w^2 \right)k_{\mbox{\tiny max}}^2 \,\alpha }
{4\, k_{\mbox{\tiny max}}^2 \,\left(w^2 - k_{\mbox{\tiny max}}^2 \right)  +
w^4\,\sinh^2{(\alpha)}}\right\}
\label{4}
\end{eqnarray}\normalsize
where we have introduced the parameter
$\alpha = \left(w^2 - k_{\mbox{\tiny max}}^2 \right)^{\frac{1}{2}}\, L$.
The concept of {\em opaque} limit appears when 
we assume that $k_{\mbox{\tiny max}}$ is independent of $L$
and then we make $\alpha$ tends to $\infty$ \cite{Jak98}.
In this case, the transit time can be rewritten as 
\small\begin{equation}
t^{^{OL}}_T = \frac{2\,m}{k_{\mbox{\tiny max}}\,\rho(k_{\mbox{\tiny max}})}.
\label{5}
\end{equation}\normalsize
In the literature, the value of $k_{\mbox{\tiny max}}$ is frequently
approximated by $k_{\0}$, the maximum of $g(k - k_{\0})$, which, in fact,
does not depend on $L$ and, under the theoretical point of view, could lead to the {\em superluminal} transmission time interpretation \cite{Jak98,Olk92,Esp03}.

It would be perfectly acceptable to consider $k_{\mbox{\tiny max}} = k_{\0}$
for the application of the stationary phase condition
if the momentum distribution $g(k - k_{\0})$ centered at $k_{\0}$ had not been
modified by any boundary condition. This is the case of the incident wave packet
before colliding with the potential barrier.  
Our criticism concern with the way of obtaining all the above results
for the the transmitted wave packet.
It has not taken into account the bounds and enhancements imposed 
by the analytical form of the transmission coefficient.

To perform the correct analysis, we should calculate the right value of
$k_{\mbox{\tiny max}}$ to be substituted in Eq.~(\ref{4}) before taking
the {\em opaque} limit.
We should be obliged to consider the relevant amplitude for the transmitted wave as 
the product of a symmetric momentum distribution $g(k - k_{\0})$ which describes the
{\em incoming} wave packet by the modulus of the transmission amplitude
$T(k, L)$ which is a crescent function of $k$.
The maximum of this product representing the transmission
modulating function would be given by the solution of the equation
\begin{eqnarray}
g(k - k_{\0}) \,\left|T(k, L)\right|\,\left[\frac{g^{\prime}(k - k_{\0})}{g(k - k_{\0})}+
\frac{\left|T(k, L)\right|^{\prime}}{\left|T(k, L)\right|}\right] &=& 0.
\label{3p42B}
\end{eqnarray}
Obviously, the peak of the   
modified momentum distribution is shifted to the right of $k_{\0}$
so that $k_{\mbox{\tiny max}}$ has to be found in the interval $]k_{\0}, w[$.
It could be numerically confirmed that $k_{\mbox{\tiny max}}$ presents an implicit
dependence on $L$ so that, by increasing the value of $L$ with respect to $a$,
the value of $k_{\mbox{\tiny max}}$
to be utilized in Eq.~(\ref{4}) would increase until $L$
reaches certain values for which
the modified momentum distribution becomes unavoidably distorted.
In this case, the relevant values of $k$ are concentrated around the upper boundary value $w$.

Due to the {\em filter effect}, the amplitude of the transmitted wave
is essentially composed by the plane wave components of the front tail of the
{\em incoming} wave packet which reaches the first barrier interface before
the peak arrival.
Meanwhile, only whether we had {\em cut} the momentum distribution {\em off} at a
value of $k$ smaller than $w$, i. e. $k \approx (1 - \delta) w$,
the {\em superluminal} interpretation 
of the transition time (\ref{5}) could be recovered.
In this case, independently of the way as $\alpha$ tends to
$\infty$, the value assumed by the transit time would be approximated by
$t^{\alpha}_{T} \approx 2 \,m / w \, \delta$ which is a finite quantity.
Such a finite value would confirm the hypothesis of {\em superluminality}.
However, the {\em cut off} of the momentum distribution
at $k \approx (1 - \delta) w$ increases the amplitude of
the tail of the incident wave and it contributes so relevantly as the peak of the incident wave to the final 
composition of the transmitted wave so that an ambiguity in the definition of the {\em arrival} time
is created.

We are particularly convinced that the use of a step-discontinuity to analyze signal transmission
in tunneling processes deserves a more careful analysis than the immediate application of the stationary phase
method since we cannot find an analytic-continuation between the {\em above} barrier case solutions and the {\em below}
barrier case solutions.
A suggestive possibility can ask for the use of the multiple peak decomposition technique developed
for the above barrier diffusion problem \cite{Ber04}.
Thus, in a similar framework, we suggest a suitable way for comprehending the conservation
of probabilities for a very particular tunneling configuration where the asymmetric aspects discussed up to now can be
totally eliminated, with the phase times being accurately calculated.    
By means of such an experimentally verifiable exercise, we shall be able to understand how the {\em filter effect} can
analytically affect the calculations of transit times in the tunneling process.

In order to recover the scattered momentum distribution symmetry conditions for correctly applying the SPM, we
assume a symmetrical colliding configuration of two wave packets traveling in opposite directions. 
By considering the same barrier represented in (\ref{2p60}), we solve the Schroedinger equation for
a plane wave component of momentum $k$ for two identical wave packets symmetrically separated from the origin $x = 0$.
At time $t = - (m L) /(2 k_{\0})$ chosen for mathematical convenience, we assume they perform a totally symmetric simultaneous collision with the potential barrier.
The wave packet reaching the left(right)-side of the barrier is thus represented by 
\small\begin{equation}
\psi^{\L(\R)}(x,t) = \int_{_{\0}}^{^{\infty}}dk \, g(k - k_{\0})\phi^{\L(\R)}(k,x)\, \exp{[- i \, E\,t]}
\end{equation}\normalsize
where, as a first approximation, we are assuming that the integral can be naturally extended from the interval $[0,w]$ to the interval $[0, \infty]$.
Its range of validity can be controlled by the choice of the width $\Delta k$ of the momentum distribution $g(k - k_{\0})$ (with $k_{\0} > 0$), i. e.
$\Delta k$ has to be bounded by the barrier high $(V_{\0})$ in order to avoid the contribution
of the above-barrier frequencies (or energies) contained in the considered wave packet (which eventually become
important as the tunnelling components are progressively damped down).

By assuming that $\phi^{\L(\R)}(k,x)$ are Schroedinger equation solutions,
when the wave packet peaks simultaneously 
reach the barrier ( at the time $t = - (m L) /(2 k_{\0})$) we can write
\small\begin{equation}
\phi^{\L(\R)}(k,x)=\left\{
\begin{array}{l l l l}
\phi^{\L(\R)}_{\1}(k,x) &=& \exp{\left[ \pm i \,k \,x\right]} + R^{\L(\R)}_{\bb}(k,L)\exp{\left[ \mp i \,k \,x\right]}&~~~~x < - L/2\, (x > L/2),\nonumber\\
\phi^{\L(\R)}_{\2}(k,x) &=& \alpha^{\L(\R)}_{\bb}(k)\exp{\left[ \mp\rho  \,x\right]} + \beta^{\L(\R)}_{\bb}(k)\exp{\left[ \pm\rho  \,x\right]}&~~~~- L/2 < x < L/2,\nonumber\\
\phi^{\L(\R)}_{\3}(k,x) &=& T^{\L(\R)}_{\bb}(k,L)\exp{ \left[\pm i \,k \,x\right]}&~~~~x > L/2 \, (x < - L/2) .
\end{array}\right.
\label{510}
\end{equation}\normalsize
where the upper(lower) sign is related to the index $L$($R$).
By assuming the conditions for the continuity of $\phi^{\L,\R}$ and their derivatives
at $x = - L/2$ and $x = L/2$, after some mathematical manipulations, we can easily obtain
\small\begin{equation}
R^{\L,\R}_{\bb}(k,L) = \exp{\left[ - i \,k \,L \right]} \left\{\frac{\exp{\left[ i \, \Theta(k,L)\right]} \left[1 - \exp{\left[ 2\,\rho(k) \,L\right]}\right]}{1 - \exp{\left[ 2\,\rho(k) \,L\right]}\exp{\left[ i\, \Theta(k,L)\right]}}\right\}
\label{511}
\end{equation}\normalsize
and
\small\begin{equation}
T^{\L,\R}_{\bb}(k,L) = \exp{\left[ - i \,k \,L \right]} \left\{\frac{\exp{\left[\rho(k) \,L\right]}\left[1- \exp{\left[ 2\, i\, \Theta(k,L)\right]}\right]}{1 - \exp{\left[ 2\,\rho(k) \,L\right]}\exp{\left[ i\, \Theta(k,L)\right]}}\right\},
\label{512}
\end{equation}\normalsize
where $\Theta(k,L)$ is given by the Eq.~(\ref{502}) and $R^{\L}_{\bb}(k,L)$ and $T^{\R}_{\bb}(k,L)$ as well as $R^{\R}_{\bb}(k,L)$ and $T^{\L}_{\bb}(k,L)$ 
are intersecting each other.
By analogy with the procedure summing amplitudes which we have adopted in the multiple peak decomposition scattering \cite{Ber04},
such a pictorial configuration obliges us to sum the intersecting amplitude of probabilities before taking their squared
modulus in order to obtain
\begin{eqnarray}
R^{\L,\R}_{\bb}(k,L)+ T^{\R,\L}_{\bb}(k,L) &=& \exp{\left[ - i \,k \,L \right]} \left\{\frac{\exp{\left[\rho(k) \,L\right]}+ \exp{\left[ i\, \Theta(k,L)\right]}}{1 + \exp{\left[ \rho(k) \,L\right]}\exp{\left[ i\, \Theta(k,L)\right]}}\right\}\nonumber\\
 &=&\exp{\left\{ - i [k \,L + \varphi(k,L)]\right\}}
\label{513}
\end{eqnarray}
with
\small\begin{equation}
\varphi(k,L) = \arctan{\left\{\frac{2\,k\,\rho(k) \, \sinh{[\rho(k)\,L]}}{w^{\2} + \left(k^{\2}-\rho^{\2}(k)\right)\cosh{[\rho(k)\,L]}}\right\}}.
\label{514}
\end{equation}\normalsize
The important information we get from the relation given by (\ref{513}) is that,
differently from the previous standard tunneling analysis, by adding the intersecting amplitudes at each side of the barrier,
we keep the original momentum distribution undistorted since $|R^{\L,\R}_{\bb}(k,L)+ T^{\R,\L}_{\bb}(k,L)|$ is equal to one.
At this point we recover the most fundamental condition for the applicability of the SPM which allows us to
accurately find the position of the peak of the reconstructed wave packet composed by reflected and transmitted
superposing components.

The phase time interpretation can be, in this case, correctly quantified in terms of the analysis of the
{\em new} phase $\varphi(k, L)$.
By applying the stationary phase condition to the recomposed wave packets, the maximal point of the
scattered amplitudes $g(k - k_{\0})|R^{\L,\R}_{\bb}(k,L)+ T^{\R,\L}_{\bb}(k,L)|$ are accurately given by $k_{\mbox{\tiny max}} = k_{\0}$
so that the traversal/reflection time or, more generically, the scattering time, results in
\small\begin{equation}
t^{^{\varphi}}_{T} =\frac{m }{k_{\0}}\left.\frac{d\varphi(k, \alpha_{(\L)})}{dk}\right|_{_{k = k_{\0}}} =
\frac{2\,m\, L}{k_{\0}\,\alpha}
\frac{w^{\2}\sinh{(\alpha)} - \alpha\,k^{\2}_{\0}}{2\,k^{\2}_{\0} - w^{\2} + w^{\2}\cosh^{\2}{(\alpha)}}
\label{515}
\end{equation}\normalsize
with $\alpha$ previously defined. 
It can be said metaphorically that the identical particles represented by both incident wave packets spend a time of the order of $ t^{^{\varphi}}_{T}$
inside the barrier before retracing their steps or tunneling.
In fact, we cannot differentiate the tunneling from the reflecting waves for such a scattering configuration.
The point is that we have introduced a possibility of improving 
the efficiency of the SPM in calculating reflecting and tunneling phase times
by studying a process where the conditions for applying the method are totally recovered,
i. e. we have demonstrated that the transmitted and reflected interfering amplitudes results
in a unimodular function which just modifies the {\em envelop} function $g(k - k_{\0})$ by an additional phase.
The previously appointed incongruities which cause the distortion of the momentum distribution $g(k - k_{\0})$ are
completely eliminated in this case.
At the same time, one could argue about the possibility of extending such a result to the standardly established
tunneling process.
We should assume that in the region inside the potential barrier, the reflecting and transmitting 
amplitudes should be summed before we compute the phase changes.
Obviously, it would result in the same phase time expression represented by (\ref{515}).
In this case, the assumption of there (not) existing interference between the momentum amplitudes of the reflected and transmitted
at the discontinuity points $x = -L/2$ and $x = L/2$ is purely arbitrary.
Consequently, it is important to reinforce the argument that such a possibility of interference leading to different phase time results
is strictly related to the idea of using (or not) the multiple peak (de)composition in the region where the potential barrier is localized.  
\begin{figure}[th]
\centerline{\psfig{file=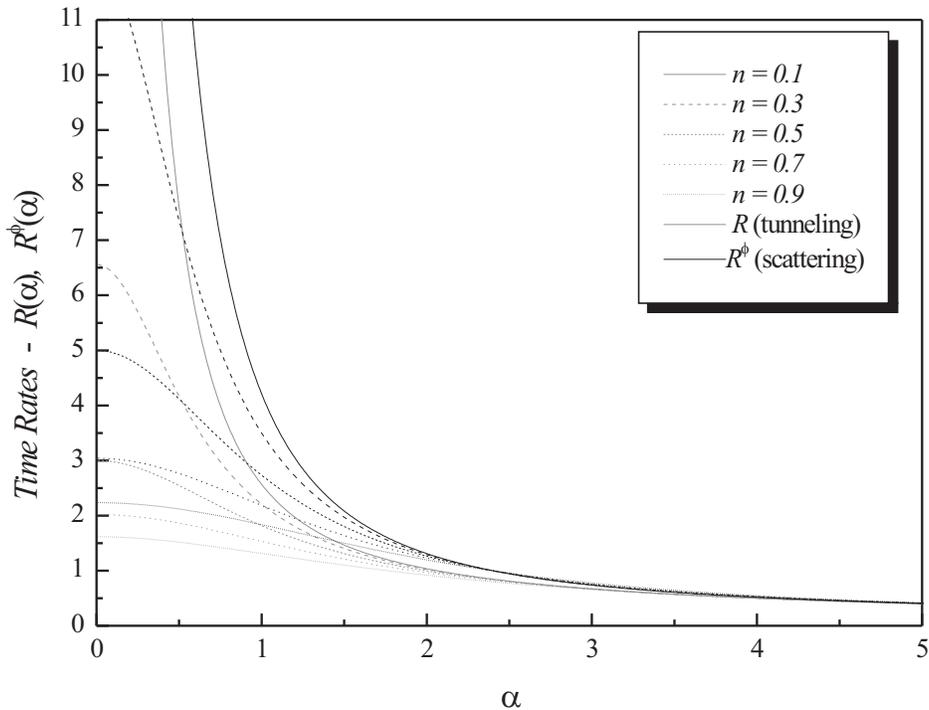,width=14cm}}
\vspace*{8pt}
\caption{Time ratios for the {\em standard} tunneling and the {\em new}
scattering process.
The ratios $R(\alpha)$ and $R^{\phi}(\alpha)$ can be understood as transmitted times in the unities of the
classical propagation time $\tau$. Both of them present the same asymptotic behavior which recover 
the theoretical possibility of a {\em superluminal} transmission in the sense that by now,  under the point of view of the analytical limitations,
the SPM can be accurately applied.
\label{fig3}}
\end{figure}
 
In order to illustrate the difference between the standard {\em tunneling} phase time $t_{T}$ and the
the alternative {\em scattering} phase time $t^{^{\varphi}}_{T}$ we introduce the new parameter $n = k^{\2}_{\mbox{\tiny max}}/w^{\2}$
and the {\em classical} traversal time $\tau = (m L) /k_{\mbox{\tiny max}}$ in order to define the ratios
\small\begin{equation}
R_{T}(\alpha) = \frac{t_{T}}{\tau}= 
\frac{2}{\alpha}\left\{
\frac{\cosh{(\alpha)}\sinh{(\alpha)} - \alpha\,n\left(2 n - 1\right)}{\left[4 n \left(1 - n\right)+\sinh^{\2}{(\alpha)}\right]}
\right\}\label{516}
\end{equation}\normalsize
and
\small\begin{equation}
R^{^{\varphi}}_{T}(\alpha) = \frac{t^{^{\varphi}}_{T}}{\tau}= 
\frac{2}{\alpha}\left\{\frac{n\, \alpha + \sinh{(\alpha)}}{2n - 1 +\cosh{(\alpha)}}
\right\}\label{517}
\end{equation}\normalsize
which are plotted in the Fig.(\ref{fig3}) for some discrete values of $n$ varying from 0 to 1,
from which we can obtain the commonest limits given by
\small\begin{equation}
\lim_{\alpha \rightarrow \infty}
{\left\{R^{^{\varphi}}_{T}(\alpha)\right\}}
= 
\lim_{\alpha \rightarrow \infty}
{\left\{R_{T}(\alpha)\right\}}
=0
\label{518}
\end{equation}\normalsize
and
\small\begin{equation}
\lim_{\alpha \rightarrow 0}
{\left\{R_{T}(\alpha)\right\}}
= 1+ \frac{1}{2 n}, ~~~~
\lim_{\alpha \rightarrow 0}
{\left\{R^{^{\varphi}}_{T}(\alpha)\right\}}
= 1+ \frac{1}{n}
\label{519}
\end{equation}\normalsize
Both of them present the same asymptotic behavior which, at first glance, recover 
the theoretical possibility of a {\em superluminal} transmission in the sense that, by now,  
the SPM can be correctly applied since the analytical limitations are accurately observed.
At this point, it is convenient to notice that the superluminal phenomena, observed in the experiments with tunnelling photons and evanescent
electromagnetic waves \cite{Nim92,Ste93,Chi98,Hay01}, has generated a lot of discussions on relativistic causality.
In fact, superluminal group velocities in connection with quantum (and classical)
tunnelings were predicted even on the basis of tunneling
time definitions more general than the simple Wigner's phase-time \cite{Wig55} (Olkhovsky {\em et al.}, for instance, discuss a simple way of understanding the problem \cite{Olk04}).
In a {\em causal} manner, it might consist in explaining the superluminal phenomena during tunnelling
as simply due to a {\em reshaping} of the pulse, with
attenuation, as already attempted (at the classical limit) \cite{Gav84}, i. e. the later parts
of an incoming pulse are preferentially attenuated, in such a way that the outcoming peak appears
shifted towards earlier times even if it is nothing but a portion of the incident pulse forward tail \cite{Ste93,Lan89}.
In particular, we do not intend to extend on the delicate question whether superluminal group-velocities can sometimes imply superluminal signalling,
a controversial subject which has been extensively explored in the literature about the tunneling effect (\cite{Olk04} and references therein).

Turning back to the scattering time analysis,
we can observe an analogy between our results and the results interpreted from the Hartman Effect (HE) analysis \cite{Har62}. 
The HE is related to the fact that for opaque potential 
barriers the mean tunnelling time does not depend on the barrier width, so that for large
barriers the effective tunnelling-velocity can become arbitrarily large,
where it was found that the tunnelling phase-time was independent of the barrier width.
It seems that the penetration time, needed
to cross a portion of a barrier, in the case of a very long barrier starts to increase again after
the plateau corresponding to infinite speed — proportionally to the distance \footnote{The validity of the HE was tested for all the other theoretical expressions proposed for the mean
tunnelling times \cite{Olk04}.}.
Our phase time dependence on the barrier width is similar to that which leads to Hartman interpretation as we can infer from Eqs.~(\ref{518}-\ref{519}).
Only when $\alpha$ tends to $0$ we have an explicit
linear time-dependence on $L$ given by
\small\begin{equation}
t^{\varphi}_T = \frac{2\, m \, L}{w} \left(1 + \frac{1}{n}\right)
\label{14}
\end{equation}\normalsize
which agree with calculations based on the simple phase-time analysis where
$t_T = \frac{2\, m \, L}{w} \left(1 + \frac{1}{2n}\right)$.
However, it is important to emphasize that the wave packets for which we compute the phase times illustrated in the Fig.(\ref{fig3})
are not {\em effectively} constructed with the same momentum distributions.
The phase $\Theta(k, L)$ appears when we treat separately the momentum amplitudes $g(k - k_{\0})\,|T(k, L)|$ and $g(k - k_{\0})|R(k, L)|$
and the other one $\varphi(k, L)$ appears when we sum the amplitudes $g(k - k_{\0})\,|T(k, L) + R(k, L)| = g(k - k_{\0})$ 
in order to obtain a symmetrical distribution which thus ``requalify'' the SPM for exactly determine the time-position of the peak of a wave packet.
In this sense, as a suggestive possibility for partially overcoming the incongruities (here appointed and quantified) which
appear when we adopt the SPM framework for obtaining tunneling phase times, we have claimed for the use of the multiple peak decomposition \cite{Ber04} technique presented in the
study of the the above barrier diffusion problem \cite{Ber04}.
We have essentially suggested a suitable way for comprehending the conservation
of probabilities for a very particular tunneling configuration where the asymmetry
presented in the previous case was eliminated, and the phase times could be accurately calculated.    
An example for which, we believe, we have provided a simple but convincing resolution.

In a more extended context, there also have been some attempts of yielding complex time delays, ultimately due to a
complex propagation constant. This has caused some controversies with denying the physical reality to an imaginary time \cite{Lan89}.
In parallel to the most sensible candidate for tunneling times \cite{Hau89,Lan94}, a phase-space
approach have been used to determine a semi-classical traversal time \cite{Xav97}.
This semi-classical method makes use of the concept of complex trajectories which, by its turn,
enables the definition of real traversal times in the complexified phase space. 
It is also commonly quoted in the context of testing different theories for temporal quantities such as arrival,
dwell and delay times \cite{Hau89,Lan94} and the asymptotic behavior at long times \cite{Jak98,Bau01}.
The configuration we have introduced in this manuscript suggests that perhaps
the idea of complexifying time should be investigated for some other scattering configurations.
We let for a subsequent analysis the suggestive possibility of investigating the validity of our approach 
when confronting with the phenomenon of one-dimensional non-resonant tunnelling through two
successive opaque potential barriers \cite{Olk02B} and with the intriguing case of multiple opaque barriers \cite{Esp03}.
Still concerning with the future theoretical perspectives, the symmetrical colliding configuration also offers the possibility of exploring some applications involving
soliton structures.
Finally, we believe all the above arguments reinforce the assertion that it is necessary to continue the search
for a generalized framework where barrier traversal times can be computed.

\begin{acknowledgments}
This work was supported by FAPESP (PD 04/13770-0).
\end{acknowledgments}


\begin{thebibliography}{99}

\bibitem{Nim92}
A. Enders, G. Nimtz, {\em J. Phys.-I} (France) {\bf 2}, 1693 (1992);\\
A. Enders, G. Nimtz, {\em J. Phys.-I} (France) {\bf 3}, 1089 (1993);\\
A. Enders, G. Nimtz, {\em Phys. Rev.} {\bf B47}, 9605 (1993);\\
A. Enders, G. Nimtz, {\em Phys. Rev.} {\bf E48}, 632 (1993).
\bibitem{Ste93}
A. M. Steinberg, P. G. Kwiat and R. Y. Chiao, {\em Phys. Rev. Lett.} {\bf 71}, 708 (1993);\\
A. Ranfagni {\it et al.}, {\em Phys. Rev.} {\bf E48}, 1453 (1993);\\
Ch. Spielman, {\it et al.}, {\em Phys. Rev. Lett.} {\bf 73}, 2308 (1994);\\ 
V. Laude and P. Tournois, {\em J. Opt. Soc. Am.} {\bf B16}, 194 (1999);\\
Ph. Balcou and L. Dutriaux, {\it et al.}, {\em Phys. Rev. Lett.} {\bf 78}, 851 (1997);
Frank {\it et al.}, {\em JETP Lett.} {\bf 12}, 605 (2002).
\bibitem{Chi98}
R. Y. Chiao, {\em Tunneling Times and Superluminality: a Tutorial}, {\em arXiv:quant-ph/}9811019.
\bibitem{Hay01}
A. Haybel and G. Nimtz, {\em Annals Phys.} {\bf 10}(Leipzig), Ed.08, 707 (2001);\\
G. Nimtz, {\em Superluminal Tunneling Devices}, {\em arXiv:hep-ph/}0204043.

\bibitem{Olk04}
V. S. Olkhovsky, E. Recami and J. Jakiel, {\em Phys. Rep.} {\bf 398}, 133 (2004).
\bibitem{Pri03}
G. Privitera {\it et al.}, {\em Rivista Nuovo Cimento} {\bf 26}, 1 (2004)
\bibitem{Lan94}
R. Landauer and Th. Martin, {\em Rev. Mod. Phys.} {\bf 66}, 217 (1994).
\bibitem{Olk92}
V. S. Olkhovsky and E. Recami, {\em Phys. Rep.} {\bf 214}, 339 (1992).
\bibitem{Hau89}
E. H. Hauge and J. A. Stovneng, {\em Rev. Mod. Phys.} {\bf 61}, 917 (1989).

\bibitem{Baz67}
A. L. Baz, {\em Sov. J. Nucl. Phys.} {\bf 4}, 182 (1967).
V. F. Rybachenko, {\em Sov. J. Nucl. Phys.} {\bf 5}, 635 (1967).
\bibitem{But83}
M. B\''{u}ttiker, {\em Phys. Rev.} {\bf B27}, 6178 (1983).
\bibitem{Hau87}
E. H. Hauge, J. P. Falck, and T. A. Fjeldly, {\em Phys. Rev.} {\bf B36}, 4203 (1987).
J. P. Falck and E.H. Hauge, {\em Phys. Rev.} {\bf B38}, 3287 (1988).
\bibitem{Fer90}
H. A. Fertig, {\em Phys. Rev. Lett.} {\bf 65}, 234 (1990).
\bibitem{Yuc92}
S. Y\''{u}cel and Eva Y. Andrei, {\em Phys. Rev.} {\bf B46}, 2448 (1992).
\bibitem{Hag92}
M. J. Hagmann, {\em Solid State Commun.} {\bf 82}, 867 (1992).
\bibitem{Bro94}
S. Brouard, R. Sala, and J. G. Muga, {\bf Phys. Rev.} {\bf A49}, 4312 (1994).
\bibitem{Olk95}
V. S. Olkhovsky, E. Recami, F. Raciti and A.K. Zaichenko, {\em J. de Physique-I} (France) {\bf 5}, 1351 (1995).
\bibitem{Jak98}
J. Jakiel, V. S. Olkhovsky, and E. Recami, {\em Phys. Lett.} {\bf A248}, 156 (1998).
\bibitem{Olk02}
V. S. Olkhovsky, E. Recami, and G. Salesi, {\em Europhys. Lett.} {\bf 57}, 879 (2002).
\bibitem{Sok87}
D. Sokolorski and L. M. Baskin, {\em Phys. Rev.} {\bf A36}, 4604 (1987).
\bibitem{Ima97}
K. Imafuku, I. Ohba, Y. Yamanaka, {\em Phys. Rev.} {\bf A56}, 1142 (1997).
\bibitem{Abo00}
M. Abolhasani and M. Golshani, {\em Phys. Rev.} {\bf A62}, 012106 (2000).
\bibitem{But82}
M. B\''{u}ttiker and R. Landauer, {\em Phys. Rev. Lett.} {\bf 49}, 1739 (1982).
\bibitem{Nim94}
G. Nimtz, A. Enders and H. Spieker,{\em J. Phys.-I} (France) {\bf 4}, 1 (1994).
\bibitem{Gar70}
C. G. B. Garret and D. E. McCumber, {\em Phys. Rev.} {\bf A01}, 305 (1970).
\bibitem{Chu82}
S. Chu and W. Wong, {\em Phys. Rev. Lett.} {\bf 48}, 738 (1982);\\
B. Segard and B. Macke, {\em Phys. Lett.} {\bf A109}, 213 (1985);\\
M. W. Mitchell and R. Y. Chiao, {\em Phys. Lett.} {\bf A230}, 122 (1997);\\
L. J. Wang, A. Kuzmich and A. Dogariu, {\em Nature} {\bf 406}, 277 (2000).

\bibitem{Boh52}
D. Bohm, {\em Quantum Theory} (Prentice-Hall, New York, 1952).
\bibitem{Wig55}
E. P. Wigner, {\em Phys. Rev.} {\bf 98}, 145 (1955).
\bibitem{PBE}
L. A. MacColl, {\em Phys. Rev.} {\bf 40}, 621 (1932).
\bibitem{Bra70}
M. H. Bramhall and C. M. Casper, {\em Am. J. Phys.} {\bf 38}, 1136 (1970).
\bibitem{Per01}
A. Péres, S. Brouard and J. G. Muga, {\em Phys. Rev.} {\bf A64}, 012710 (2001).
\bibitem{Gav84}
B. Gaveau {\it et al.}, {\em Phys. Rev. Lett.} {\bf 53}, 419 (1984).
\bibitem{Lan89}
R. Landauer, {\em Nature} {\bf 341}, 567 (1989).
\bibitem{Har62}
T. E. Hartman, {\em J. Appl. Phys.} {\bf 33}, 3427 (1962).
\bibitem{Ber04}
A. E. Bernardini, S. De Leo and P. P. Rotelli, {\em Mod. Phys. Lett} {\bf A19}, 2717 (2004).
\bibitem{Coh77}
C. Cohen-Tannoudji, B. Diu and F. Lal\"oe, {\em Quantum Mechanics}
(John Wiley \& Sons,  Paris, 1977), pag.\,81.
\bibitem{Xav97}
A. L. Xavier and M. A. M. de Aguiar, {\em Phys. Rev. Lett.} {\bf 79}, 3323 (1997).
\bibitem{Bau01}
A. D. Baute, I. L. Egusquiza and J. G. Muga, {\em Phys. Rev.} {\bf A64}, 012501 (2001).
\bibitem{Olk02B}
V. S. Olkhovsky, E. Recami and G. Salesi, {\em Europhys. Lett.} {\bf 57}, 879 (2002).
\bibitem{Esp03}
S. Esposito, {\em Phys. Rev.} {\bf E67}, 016609 (2003).
\end{thebibliography}
\end{document}